\documentclass[graybox]{svmult}

\usepackage{type1cm}  
\usepackage{subcaption}      
\usepackage{makeidx}         
\usepackage{graphicx} 
\usepackage{graphicx}   
\usepackage{float}    
\usepackage{multicol}        
\usepackage[bottom]{footmisc}

\usepackage{newtxtext}       %
\usepackage{newtxmath}       

\usepackage{epsfig}
\usepackage{graphicx}
\usepackage[latin1]{inputenc}
\usepackage[T1]{fontenc}
\usepackage{mathptmx}
\usepackage{url}
\usepackage{amsmath}
\usepackage{amssymb}
\usepackage{hyperref}
\usepackage{lineno}

\newcommand{\sqrtSnn}{\ensuremath{\sqrt{s_{\mathrm{NN}}}}}
\newcommand{\pPb}{\ensuremath{\mbox{p--Pb}}}
\newcommand{\jpsi}{\ensuremath{{\rm J}\kern-0.02em/\kern-0.05em\psi}}

\usepackage{blindtext}

\makeindex             

\begin{document}

\title*{Inclusive $\Upsilon$ production in $\pPb$ collisions at  $\sqrtSnn$ = 8.16 TeV with ALICE at the LHC}
\author{Wadut $ \rm Shaikh^{*}$ (for the ALICE collaboration)}
\institute{Saha Institute of Nuclear Physics, Kolkata-700064, India, \email{wadut.shaikh@cern.ch}
}
%
%
\maketitle

\vspace{-3.2cm}

\abstract{Quarkonium, a bound state of a heavy quark and an anti-quark, is an important probe to study the properties of the Quark--Gluon Plasma (QGP) created in heavy-ion collisions at LHC energies. Color screening and regeneration phenomena influence the production of quarkonium due to the presence of the QGP. However the regeneration effect is small in case of $\Upsilon$ (bound state of b and $\bar{\rm b}$ quarks). Cold Nuclear Matter (CNM) effects, which include shadowing or gluon saturation and energy loss, can also lead to a modification of quarkonium production. In order to disentangle CNM effects from the ``hot'' QGP effects, quarkonium production is studied in p--Pb collisions in which QGP formation is not expected. ALICE has measured the $\Upsilon$ production in p--Pb collisions at $\sqrtSnn$ = 5.02 TeV at backward ($-$4.46 $<y_{\rm cms}<$ $-$2.96) and forward (2.03 $< y_{\rm cms}<$  3.53) rapidity down to zero transverse momentum. At forward rapidity, a suppression of $\Upsilon$(1S) production in p--Pb collisions is observed compared to the binary-scaled yield in pp collisions in the same kinematic domain while at backward rapidity no significant suppression or enhancement is found within the experimental uncertainties. In 2016, the LHC delivered p--Pb collisions at $\sqrt{s_{\rm NN}}$ = 8.16 TeV with higher integrated luminosity compared to the data collected at $\sqrt{s_{\rm NN}}$ = 5.02 TeV, which allowed a more detailed study of the bottomonium production in p--Pb collisions. We  report on the inclusive $\Upsilon$(1S) production as a function of rapidity, transverse momentum ($p_{\rm T}$) and centrality of the collision and compare the results with those obtained at $\sqrt{s_{\rm NN}}$ = 5.02 TeV. Theoretical model predictions as a function of $y_{\rm cms}$ and $p_{\rm T}$ are also discussed. The results of $\Upsilon$(2S) suppression integrated over  $y_{\rm cms}$, $p_{\rm T}$ and centrality are also reported and compared to the corresponding $\Upsilon$(1S) measurement. }
\vspace{-0.3cm}
\section{Introduction}
\vspace{-0.4cm}
Quarkonia are well known probes to study the properties of the deconfined medium, called Quark--Gluon Plasma (QGP), created in ultra relativistic heavy-ion collisions. The modification of quarkonium production in heavy-ion collisions with respect to the binary-scaled yield in pp collisions at the LHC is explained by suppression of quarkonia via color screening mechanism ~\cite{Matsui} and (re)generation of quarkonia. In the color screening mechanism, color charges present in the deconfined medium screen the binding potential between the $q$ and $\bar{q}$ quarks, leading to a temperature-dependent melting of the quarkonium states according to their binding energies. For bottomonia, (re)generation effects are expected to be negligible due to the small number of produced $b$ quarks. Cold nuclear matter effects (shadowing, parton energy loss, interaction with hadronic medium) which are not related to the deconfined medium may also lead to a modification of quarkonium production. In order to disentangle the CNM effects from the hot nuclear matter effects, quarkonium production is studied in p--Pb collisions in which the QGP is not expected to be formed.



The inclusive $\Upsilon$ production in p--Pb collisions  at  $\sqrt{s_{\rm NN}}$ = 8.16 TeV is studied with the ALICE detector at the CERN LHC. The measurement
has been performed reconstructing $\Upsilon$(1S) and $\Upsilon$(2S) mesons via their dimuon decay channel, in the
forward (proton-going direction) and backward (lead-going direction) rapidity ranges down to zero transverse momentum with integrated luminosities of 8.4$\pm$0.2 n$\rm b^{-1}$ and 12.8$\pm$0.3 n$\rm b^{-1}$, respectively.

\vspace{-0.3cm}
\section{Results}
\vspace{-0.25cm}
The cold nuclear matter effects in p--Pb collisions can be quantified through the nuclear modification factor defined as \\

\begin{equation}
R_{\rm pPb} = \frac{N_{\rm \Upsilon}}{\langle T_{\rm pPb}\rangle \times (A\times\varepsilon) \times N_{\rm MB}\times{\rm BR_{\rm \Upsilon\rightarrow\mu^{+}\mu^{-}}}\times\sigma^{\rm pp}_{\rm \Upsilon}}
 \label{eq:1}
\end{equation}

where:
\begin{itemize}
\item[{ $\bullet$}] $N_{\rm \Upsilon}$ is the number of $\Upsilon$ in a given $y_{\rm cms}$, $p_{\rm T}$  or centrality bin.
\item[{ $\bullet$}] $\langle T_{\rm pPb}\rangle$ is the centrality-dependent average nuclear overlap function.
\item[{ $\bullet$}] $A\times\epsilon$ is the product of the detector acceptance and the reconstruction efficiency.
\item[{ $\bullet$}] $N_{\rm MB}$ is the number of collected minimum-bias events.
\item[{ $\bullet$}] BR$_{\rm \Upsilon\rightarrow\mu^{+}\mu^{-}}$ is the branching ratio of $\Upsilon$ in the dimuon decay channel (BR$_{\rm \Upsilon(1S)\rightarrow\mu^{+}\mu^{-}}$=
2.48$\pm$0.05$\%$, BR$_{\rm \Upsilon(2S)\rightarrow\mu^{+}\mu^{-}}$=1.93$\pm$0.17$\%$)  ~\cite{pdg}.
\item[{ $\bullet$}] $\sigma^{\rm pp}_{\rm \Upsilon}$ is the inclusive $\Upsilon$ production cross section for pp collisions at the same energy, $y_{\rm cms}$ and $p_{\rm T}$  interval as for p--Pb collisions. These values have been obtained by means of an interpolation procedure based on the LHCb pp $\Upsilon$ measurements ~\cite{LHCb_pp}. 
\end{itemize} 

\begin{figure}[!h]
\hspace{-3cm}
	\centering
	\begin{minipage}[t]{4cm}
		\centering
		\includegraphics[scale=0.3]{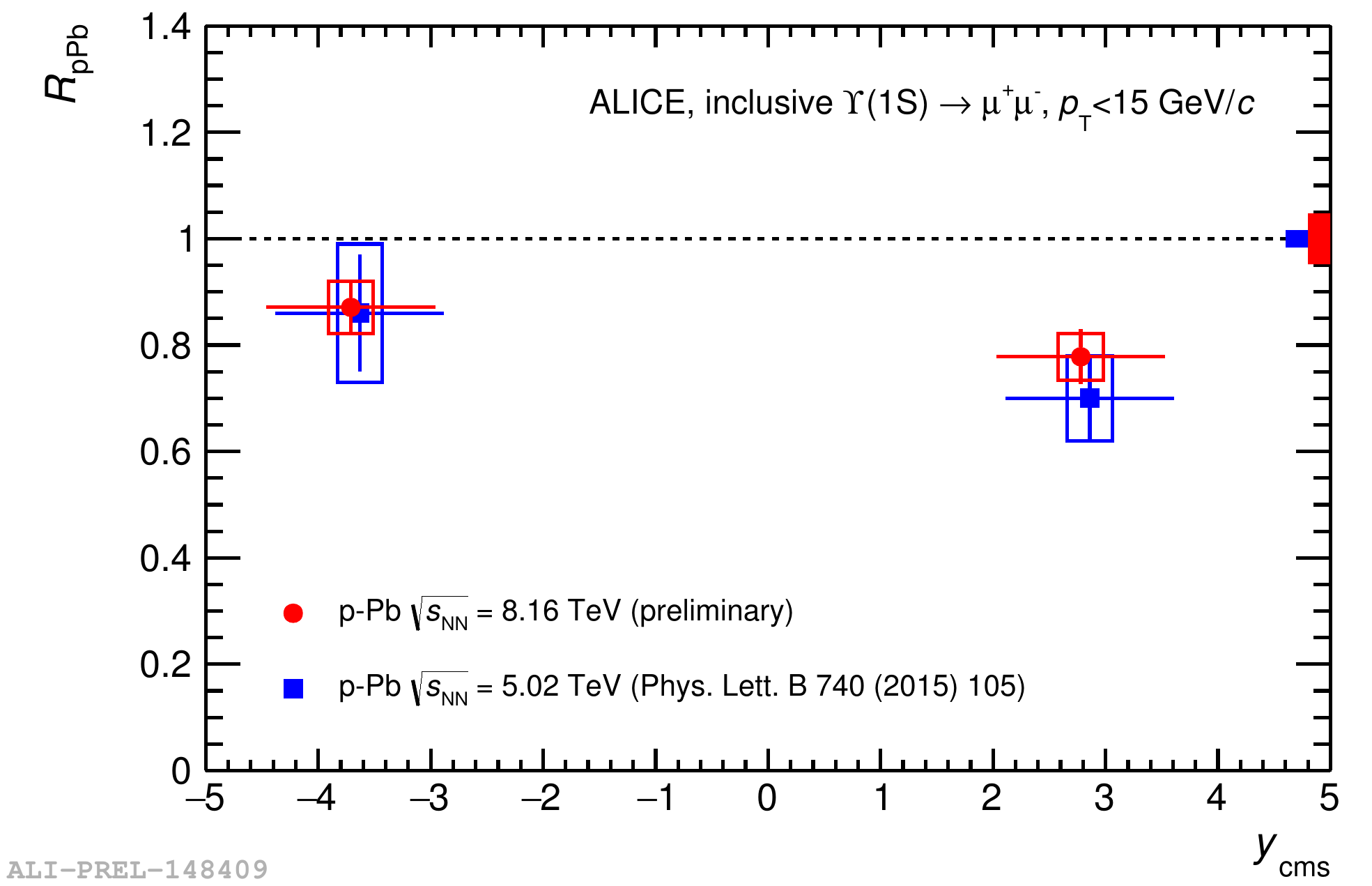}
	
	\end{minipage}
	\hspace{2cm}
	\begin{minipage}[t]{4cm}
		\centering
		\includegraphics[scale=0.3]{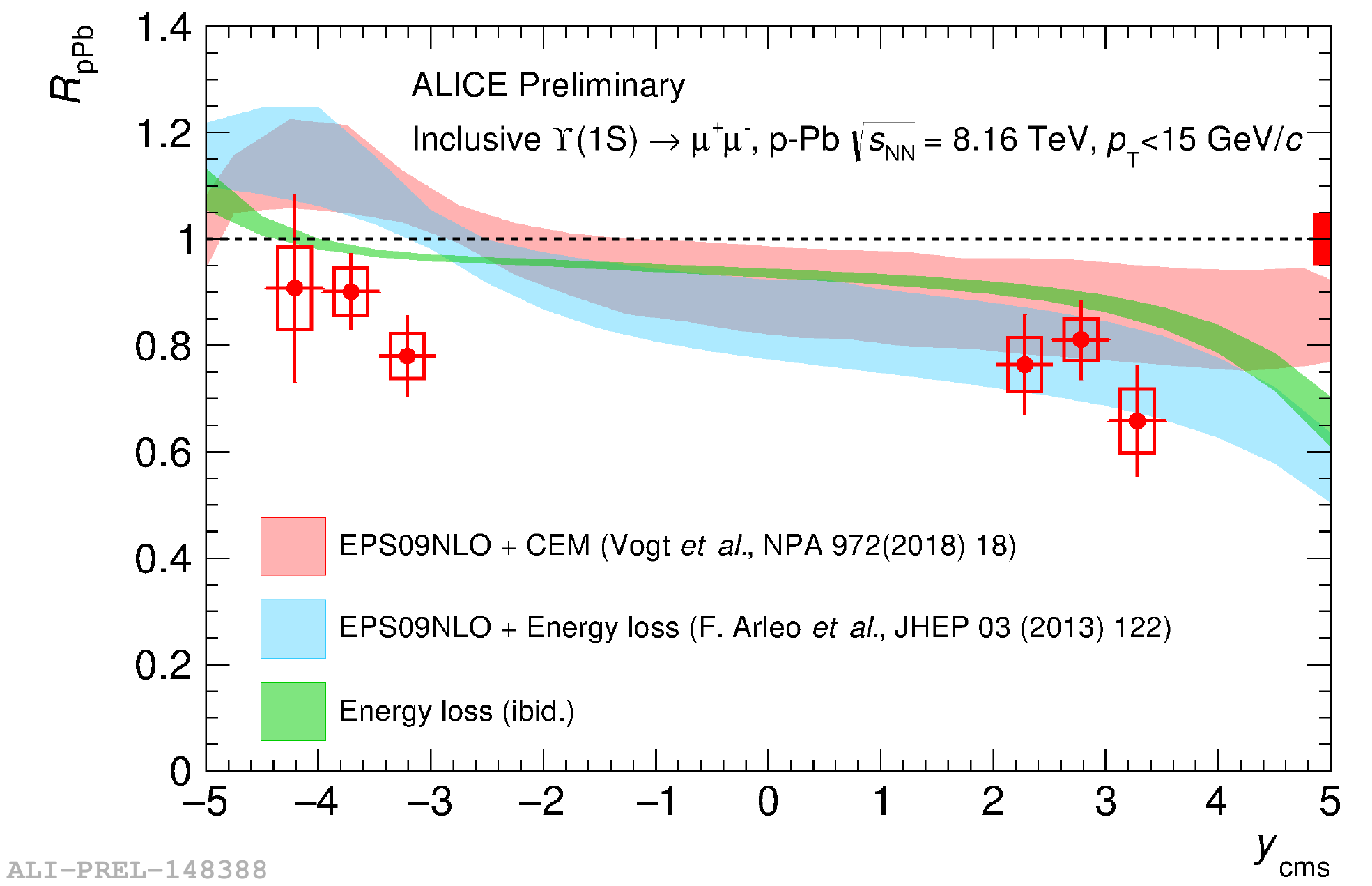}
	
	\end{minipage}
		\caption{ Integrated $R_{\rm pPb} $ of  $\Upsilon$(1S) in p--Pb collisions at $\sqrt{s_{\rm NN}}$= 8.16 TeV compared to 5.02 TeV (left). $\Upsilon$(1S) $R_{\rm pPb} $ as function of $y_{\rm cms}$ at $\sqrt{s_{\rm NN}}$= 8.16 TeV and the values are  compared to theoretical calculations based on different CNM effects (right).}
		\label{fig:RpA_Y}
	
	\end{figure}

The inclusive $\Upsilon$(1S) nuclear modification factor at $\sqrt{s_{\rm NN}}$ = 8.16 TeV for the two studied beam configurations is shown in Fig.~\ref{fig:RpA_Y} (left panel). In this figure, as well as all the other figures, the vertical error bars represent the statistical uncertainties and the open boxes represent the uncorrelated systematic uncertainties. The full boxes around $R_{\rm pPb} $ = 1 show the size of the correlated systematic uncertainties. The measured $R_{\rm pPb}$ values indicate $\Upsilon$(1S) suppression both at forward and backward rapidity. The $R_{\rm pPb}$ values are also compared with previous ALICE measurements at $\sqrt{s_{\rm NN}}$ = 5.02 TeV ~\cite{UpsipPb5TeV} and results are compatible within the uncertainties.

 The rapidity dependence of the  $R_{\rm pPb}$ is shown in Fig.~\ref{fig:RpA_Y} (right panel). The suppression already observed in the integrated case is confirmed, the size of the uncertainties does not allow us to draw more detailed conclusions on the rapidity dependence.

\begin{figure}[!h]
\hspace{-3cm}
	\centering
	\begin{minipage}[t]{4cm}
		\centering
		\includegraphics[scale=0.3]{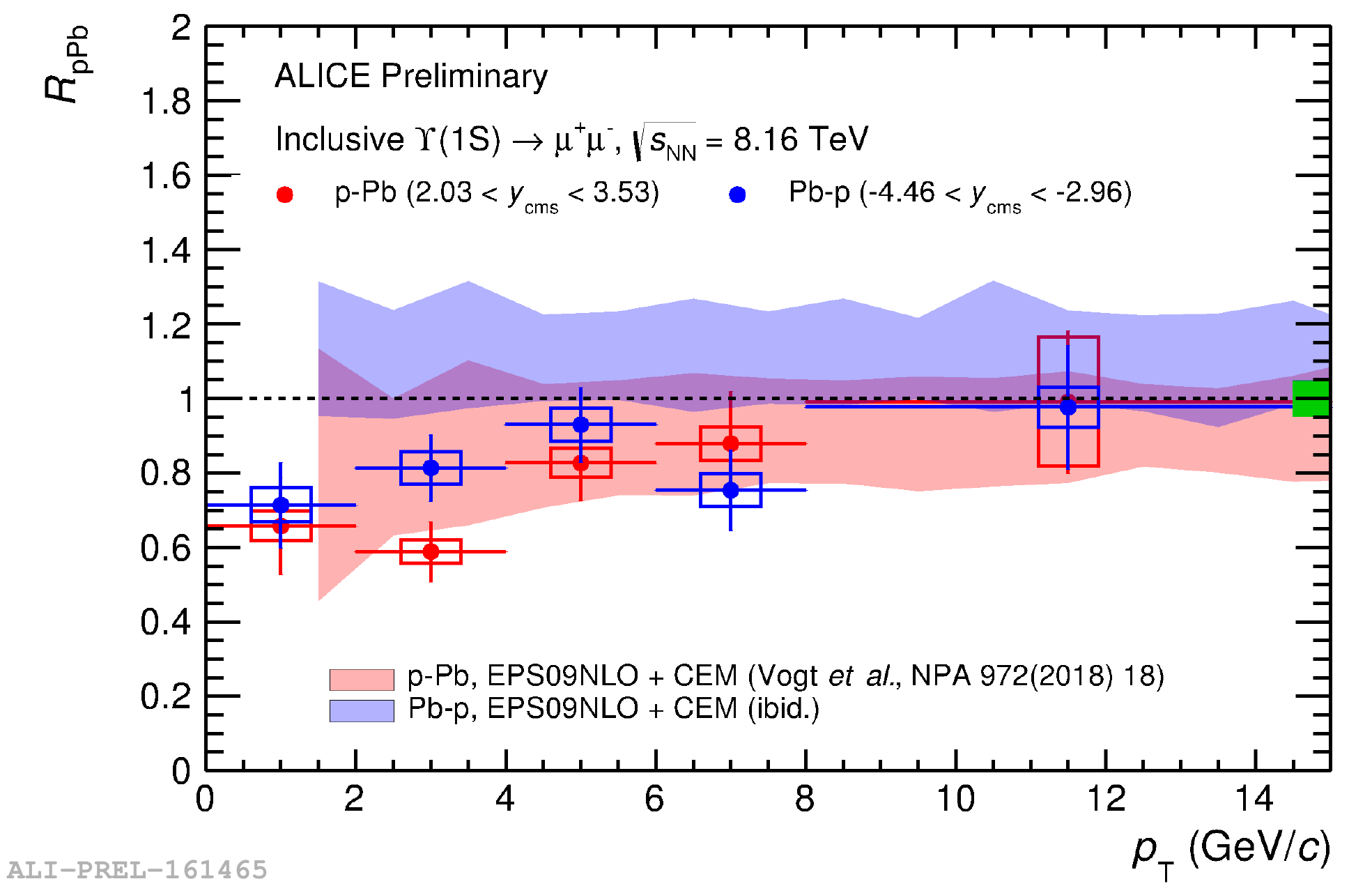}
	
	\end{minipage}
	\hspace{2cm}
	\begin{minipage}[t]{4cm}
		\centering
		\includegraphics[scale=0.3]{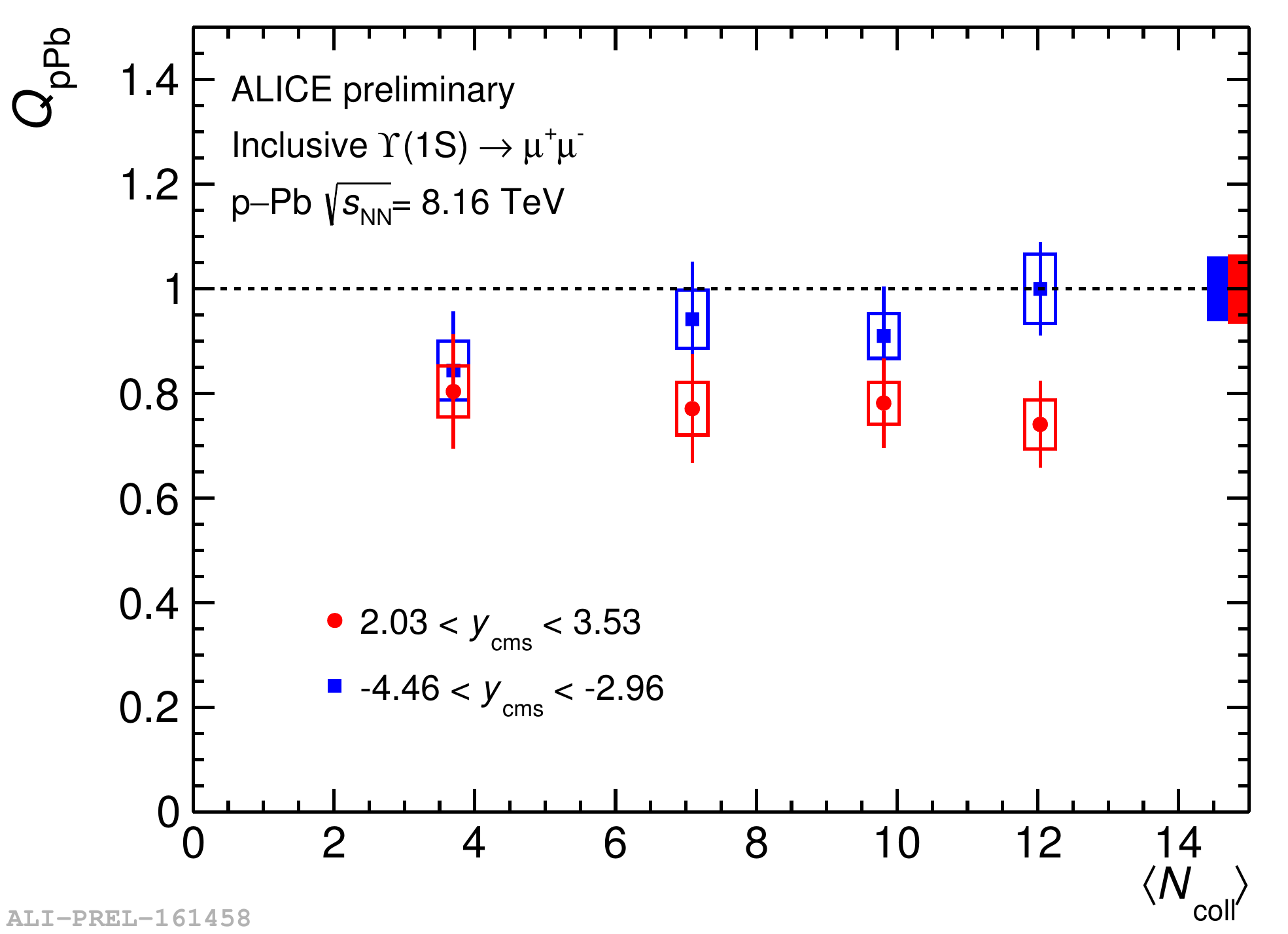}
	
	\end{minipage}
		\caption{ $\Upsilon$(1S) $R_{\rm pPb} $ as function of $p_{\rm T}$ at $\sqrt{s_{\rm NN}}$= 8.16 TeV compared to theoretical calculations based on different CNM effects (left). $\Upsilon$(1S) $Q_{\rm pPb} $ as a function of average number of binary nucleon--nucleon collisions ($\langle N_{\rm coll}\rangle$) at $\sqrt{s_{\rm NN}}$= 8.16 TeV (right).}
		\label{fig:RpA_pT_Cent}
	
	\end{figure}

The $p_{\rm T}$ dependence of the $\Upsilon$(1S) $R_{\rm pPb} $ is shown in  Fig.~\ref{fig:RpA_pT_Cent} (left panel).
A suppression of low-$p_{\rm T}$ $\Upsilon$(1S) is observed both at forward and backward rapidities.

	
	The rapidity and $p_{\rm T}$ dependences of the  $R_{\rm pPb}$ are compared to a next-to-leading order (NLO) CEM calculation using the EPS09 parameterization of the nuclear modification of the gluon PDF~\cite{Rvogt,Albacete} and to a parton energy loss calculation~\cite{Arleo} with and without EPS09 gluon shadowing at NLO. The shadowing and energy loss calculations describe the $p_{\rm T}$ and rapidity dependent results at forward rapidity within uncertainties while they overestimate the data at backward rapidity.\\

	The $\Upsilon$(1S) nuclear modification factor has been also studied as a function of the collision centrality. $Q_{\rm pPb}$ is used instead of $R_{\rm pPb}$ due to a possible bias in the centrality determination~\cite{jpsipPb5TeVcent,jpsipPb8TeVcent} and it is defined the same way as $R_{\rm pPb}$ in equation ~\ref{eq:1} . The centrality dependence of $\Upsilon$(1S) $Q_{\rm pPb}$ is shown in Fig.~\ref{fig:RpA_pT_Cent} (right panel). The $Q_{\rm pPb}$ is found to be independent of the centrality within uncertainties ~\cite{UpsipPb8TeV}.\\
	
\begin{figure}[!h]
\hspace{-3cm}
	\centering

		\includegraphics[scale=0.3]{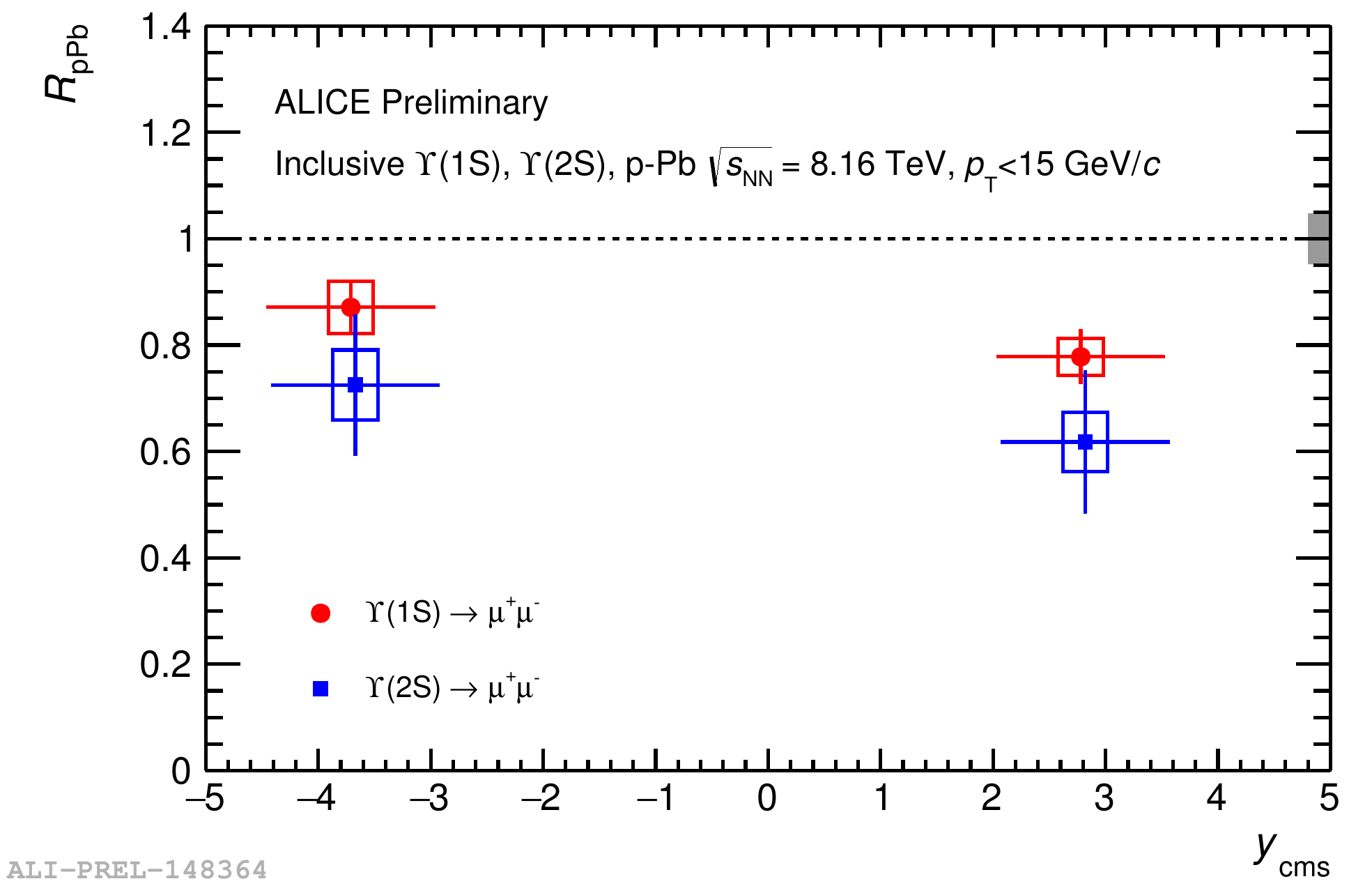}	
		
		\caption{ Integrated $R_{\rm pPb} $ of  $\Upsilon$(1S) and $\Upsilon$(2S)  in p--Pb collisions at $\sqrt{s_{\rm NN}}$= 8.16 TeV.}
			\label{fig:2S1S}
		\end{figure}

	 The lower $\Upsilon$(2S) yield does not allow differential studies, hence only results integrated over $y$, $p_{\rm T}$ and centrality are presented in Fig.~\ref{fig:2S1S}

	\vspace{-0.4cm}

The two resonances show a similar suppression, slightly larger for $\Upsilon$(2S)~\cite{UpsipPb8TeV}. The difference in the $R_{\rm pPb}$ of the $\Upsilon$(2S) and $\Upsilon$(1S) amounts to 1$\sigma$ at forward-$y$ and 0.9$\sigma$ at backward-$y$. The CMS~\cite{CMS}, ATLAS~\cite{ATLAS} and LHCb~\cite{LHCb} collaborations also observed a larger suppression for the $\Upsilon$(2S) than for the $\Upsilon$(1S).

\vspace{-0.40cm}
\section{Conclusions}
\vspace{-0.2cm}
We have presented new results on the  $\Upsilon$(1S) and  $\Upsilon$(2S) nuclear modification factor in p--Pb collisions at $\sqrt{s_{\rm NN}}$ = 8.16 TeV, measured by ALICE. The $R_{\rm pPb}$ of $\Upsilon$(1S) is similar at forward and backward rapidities with a hint for a stronger suppression at low $p_{\rm T}$. In both rapidity intervals there is no evidence for a centrality dependence of the $\Upsilon$(1S) $Q_{\rm pPb} $. Models based on nuclear shadowing and coherent parton energy loss fairly describe the data at forward rapidity, while they tend to overestimate the $R_{\rm pPb}$  at backward rapidity. The $\Upsilon$(2S) and $\Upsilon$(1S) suppressions are compatible within 1$\sigma$.
\vspace{-0.4cm}
%
%


\begin{thebibliography}{8}


\bibitem{Matsui} T. ~Matsui and H.~Satz, Phys.\ Lett.\ B 178 (1986) 416

\bibitem{UpsipPb5TeV} ALICE Collaboration, Phys.\ Lett.\ B 740   (2015) 105


\bibitem{pdg} Particle Data Group Collaboration, Chin. Phys. C40  (2016) 100001.
\bibitem{LHCb_pp} LHCb Collaboration, EPJC 74 (2014) 2835, JHEP 11 (2015) 103.
\bibitem{UpsipPb8TeV} ALICE Collaboration,  CERN-ALICE-PUBLIC-2018-008 

\bibitem{jpsipPb5TeVcent} ALICE Collaboration, JHEP 11 (2015) 127
\bibitem{jpsipPb8TeVcent} ALICE Collaboration, CERN-ALICE-PUBLIC-2017-007

\bibitem{Rvogt} R. Vogt, Phys. Rev. C92 (2015) 034909
\bibitem{Albacete}J. L. Albacete et al., Nucl. Phys. A972 (2018) 18
\bibitem{Arleo} {F. Arleo {\em et~al.}} JHEP 03 (2013) 122



\bibitem{CMS} CMS Collaboration, JHEP 04 (2014) 103
\bibitem{ATLAS} ATLAS Collaboration, EPJC 78 (2018) 1717
\bibitem{LHCb} LHCb Collaboration, JHEP 11 (2018) 194


 







\end{thebibliography}
%

\small{

}

\vspace{-0.8cm}
\end{document}